\documentclass[twocolumn,nofootinbib,superscriptaddress]{revtex4-1}
 \pdfoutput=1
 \usepackage{latexsym,epsfig,amssymb, amsmath, nicefrac}

 \usepackage{soul}
 \usepackage{color}
 \usepackage[makeroom]{cancel}
 \usepackage{latexsym}
 \usepackage{epsf}
 \usepackage{amssymb}
 \usepackage{graphicx}
 \usepackage{amsmath}
 \usepackage{amsmath,amssymb,amsthm}
 \usepackage{verbatim}
 \usepackage{hyperref}
 \renewcommand{\d}{\textrm{d}}
 \newcommand{\e}{\textrm{e}}

 \newcommand{\cD}{\mathcal{D}}
 \newcommand{\cN}{\mathcal{N}}

 \def\tb0{\tilde{\beta}_0}
 {\def\b0{\beta_0}

 	\def\bi{\begin{itemize}}
 		\def\ei{\end{itemize}}
 	\def\be{\begin{equation}}
 	\def\ee{\end{equation}}
 	\newcommand{\bea}{\begin{eqnarray}}
 	\newcommand{\eea}{\end{eqnarray}}

 	\def\a{\alpha}
 	\def\b{\beta}
 	\def\g{\gamma}
 	
 	\def\d{\delta}
 	
 	\def\e{\epsilon}
 	
 	\def\f{\phi}
 	\def\vf{\varphi}

 	\def\l{\lambda}
 	\def\L{\Lambda}
 	\def\m{\mu}
 	\def\n{\nu}
 	\def\r{\rho}
 	\def\s{\sigma}

 	\def\o{\omega}
 	
 	\def\x{\xi}
 	\def\cD{\mathcal{D}}
 	\def\cL{\mathcal{L}}
 	
 	\def\Kahler{K\"{a}hler~}

 	\def\nn{\nonumber}

\begin{document}
 		
 		\title{Broken Scale Invariance, $\a$-Attractors and Vector Impurity}
 		
 		\author{\"{O}zg\"{u}r Akarsu}
 		\email{akarsuo@itu.edu.tr}
 		\author{Sibel Boran}
 		\email{borans@itu.edu.tr}
 		\author{Emre Onur Kahya}
 		\email{eokahya@itu.edu.tr}
 		\author{Ne\c{s}e \"{O}zdemir}
 		\email{nozdemir@itu.edu.tr}
 		\author{Mehmet Ozkan}
 		\email{ozkanmehm@itu.edu.tr}
 		\affiliation{Department of Physics, Istanbul Technical University, Maslak, 34469 Istanbul, Turkey}

 		\date{\today}
 		
 		\begin{abstract}
 			
 			We show that if the $\alpha$-attractor model is realized by the spontaneous breaking of the scale symmetry, then the stability and the dynamics of the vector field that gauges the scale symmetry can severely constrain the $\alpha$-parameter as $5/6 < \alpha < 1$ restricting the inflationary predictions in a very tiny region in the $n_s - r$ plane that are in great agreement with the latest Planck data. Although the different values of $\alpha$ do not make a tangible difference for $n_s$ and $r$, they provide radically different scenarios for the post-inflationary dynamics which determines the standard BBN processes and the large scale isotropy of the universe. 
 			
 		\end{abstract}
 		
 		\maketitle

 		%\pacs{98.80.Cq, 98.80.Hw}

 		%%%%%%%%%%%%%%%%%
 		\section{Introduction}
 		\label{intro}
 		%%%%%%%%%%%%%%%%%
 		
 		Inflation is the major candidate to explain the spatial flatness, large scale isotropy and homogeneity of the observed universe, and is one of the most active research areas in high energy physics since early 1980s \cite{Starobinsky:1980te,aftstr1,aftstr2,aftstr3,oldinf1,oldinf2,oldinf3,oldinf4,newinf1,newinf2,chaotic}. The realization of the inflationary paradigm requires a scalar degree of freedom, which can be achieved in a number of ways, including a minimally coupled scalar field theory, $f(R)$ theories, and non-minimal coupling of scalar field to gravity (for reviews, see references \cite{Baumann:2009ds,Linde:2014nna,Nojiri:2010wj,Capozziello:2011et,Clifton:2011jh,DeFelice:2010aj}). Among these, the minimally coupled scalar with a potential $V(\f)$,
 		\bea
 		S &=& \int {\rm d}^4 x \sqrt{-g} \, \left(\frac{M_{\rm pl}^2}{2} R - \frac12 \partial_\m \f \partial^\m \f - V(\f) \right) \,,
 		\eea
 		contains the most of the inflationary scenarios as the others can be written as a minimally coupled theory by means of a field redefinition \cite{DeFelice:2010aj}. 
 		
 		The form of the scalar potential in a minimally coupled scenario is completely unconstrained as long as it satisfies the Poincar\'e invariance, and the slow-roll conditions. A list of possible scalar potentials can be found in \cite{Martin:2013tda}. Despite the diversity of the scalar potentials, and different ways to introduce the scalar degree of freedom to gravity, a wide variety of inflationary models were found to give rise to the same predictions for the inflationary observables \cite{Starobinsky:1980te,Salopek:1988qh,Bezrukov08,Ellis:2013a,Ellis:2013b,Kallosh:2013wya,Kallosh:2013xya,Kallosh:2013hoa,Ferrara:2013,Kallosh:2013tua,Kallosh:2013yoa,Kehagias:2013}, which are in great agreement with the latest Planck data \cite{Ade:2015lrj}. A systematic approach to investigate the root cause of this fact could be to introduce more symmetries to the theory, as more symmetry implies more restriction on the form of the scalar potential \cite{Kallosh:2013hoa}. Then, the spontaneous breaking of such symmetries can provide a novel framework to understand why many different inflationary scenarios give rise to the same observables.
 		
 		The usual dictum that the local symmetries enrich the field content of a theory by the addition of gauge fields can, in principle, demotivate one to work with extended local spacetime symmetries in the context of inflationary cosmology. This is mainly due to the fact that a massive $p$-form field, which arises after gauge fixing, can spoil the spatially homogeneous and isotropic large scale structure of the universe. One particular exception to this rule is the so-called conformal gravity
 		\bea
 		S &=& \int \, {\rm d}^4 x \, \sqrt{-g} \, C_{\m\n\r\s} C^{\m\n\r\s} \,,
 		\eea 
 		where $C_{\m\n\r\s}$ is the Weyl tensor. This model is invariant under the local conformal transformation $\widehat{g}_{\m\n} = e^{-2\Lambda_D (x)} g_{\m\n}$ without introducing gauge fields that gauge the conformal symmetries. Another example is the conformally coupled scalar-tensor theory
 		\bea
 		S &=& \int {\rm d}^4 x \, \sqrt{-g} \Big( \frac{1}{12} \f^2 R + \frac12 \partial_\m \f \partial^\m \f \Big) \,,
 		\label{ConformalScalarTensor}
 		\eea
 		which is invariant under the local conformal transformations  $\widehat{g}_{\m\n} = e^{-2\Lambda_D (x)} g_{\m\n}$ and $\widehat{\f} = e^{\L_D (x)} \f$. In recent years, the conformally coupled scalar-tensor theory with two scalar fields has been widely used in inflationary cosmology as the backbone of the superconformal $\a$-attractor program \cite{Kallosh:2013yoa}. When the local superconformal symmetries are fixed, the bosonic part of the $\a$-attractors are given by the Einstein-Hilbert action with a canonical scalar $\vf$, and a scalar potential
 		\bea
 		V(\vf) &=& f(\tanh \frac{\vf}{\sqrt{6\a} M_{\rm pl}}) \,,
 		\eea
 		where $\a$ is inversely proportional to the curvature of the \Kahler manifold \cite{Kallosh:2013yoa}. For $\a \lesssim 1$, and $N \gg 1$, with $N$ being the number of e-foldings, the model provides a universal attractor regime \cite{Kallosh:2013hoa} with predictions for the spectral index $n_s$ and the tensor to scalar ratio $r$ \cite{Kallosh:2013yoa},
 		\bea
 		n_s = 1 - \frac2N, \qquad  r = \frac{12 \a}{N^2} \,,
 		\label{Universal}
 		\eea
 		which reduces to the predictions of the aforementioned inflationary models for $\a=1$, and are favored by the Planck data \cite{Ade:2015lrj}.
 		
 		The $\a$-attractors lead to inflationary predictions that perfectly fits to the latest Planck data \cite{Ade:2015lrj} for a wide range of $\a$. In the context of superconformal $\a$-attractors, the $\a$ parameter is a measure of the curvature of the inflaton K\"ahler manifold \cite{Kallosh:2013yoa}. The $\a$ parameter also appears in other contexts such as the auxiliary vector modification \cite{Ozkan:2015iva}, string field theory approach \cite{Koshelev:2016vhi}, spacetime with vector distortion \cite{Jimenez:2015fva,Jimenez:2016wbq}, and scale invariant gravity  \cite{Ozkan:2015kma}. Especially, in the context of scale invariance, the parameter $\a$ appears naturally as a quantity that measures the deviation from conformal symmetry while preserving the scale symmetry of the theory.
 		
 		As mentioned, while the conformal gravity does not add extra gauge fields to the field content of the theory, the scale invariance does add a vector field that gauges the scale symmetry. Therefore, the price to pay for relaxing the conformal symmetry is the possibility of a large scale anisotropy sourced by the vector field. In this paper, our purpose is to investigate the features of the vector field that appears in the local scale invariant setting of $\a$-attractors. As opposed to the usual expectation that relaxing the conformal symmetry should loosen any possible constraint on $\a$ that the conformal symmetry might dictate, we find that the stability of the theory in the scale symmetric setting can severely constrain $\a$, namely, $5/6 < \a < 1$ for the model we introduced in this paper. Thus, although the scalar potential for the conformal and the scale invariant $\a$-attractors are precisely the same, the dynamics of the vector field that appears in the scale invariant setting can be a constraining factor in the inflationary dynamics.
 		
 		The paper is organised as follows. In Section \ref{WeylAndConformal}, we review the basics of the local Poincar\'e, scale and special conformal invariance. In Section \ref{ConservedCharge}, we explicitly calculate the conserved current for the scale transformations and show that it does not vanish unless the scale invariance is enhanced to the full conformal invariance. In Section \ref{AlphaAttractors}, we discuss the gauge fixing of the local scale invariant theory, and the $\a$-attractors. In Section \ref{VectorFieldStability}, we discuss the consequences of the stability and dynamics of the vector field in the inflationary and post-inflationary eras. Section \ref{Conclusions} is devoted to the conclusions.
 		
 		%%%%%%%%%%%%%%%%%
 		\section{Basics of Local Scale and Conformal Invariance}
 		\label{WeylAndConformal}
 		%%%%%%%%%%%%%%%%%
 		To set the stage, we first review the consequences of the local scale and conformal invariance in gravity theories. Our treatment follows the discussions presented in \cite{Freedman:2012zz}. Therefore, we refer the reader interested in a more detailed discussion on conformal (or scale invariant) construction of (super)gravity models to \cite{Freedman:2012zz}.  
 		
 		Considering gravity as a gauge theory is based on the Poincar\'e group with the generators
 		\bea
 		P_a\,, \quad M_{ab} \,,
 		\label{PoincareGenerators}
 		\eea
 		and the gauge fields for these generators
 		\bea
 		h_\m{}^A \equiv \{e_\m{}^a\,, \o_\m{}^{ab} \} \,,
 		\label{GaugeFieldsPoincare}
 		\eea
 		where $\m,\n\,\ldots$ are world vector indices and $a,b,\ldots$ are Lorentz indices. Using the structure constants of the four dimensional Poincar\'e algebra $f_{BC}{}^A$ and the basic rules
 		\bea
 		\d h_\m{}^A &=& \partial_\m \e^A + f_{BC}{}^A h_\m^B \e^C \,,\nn\\
 		R_{\m\n}{}^A &=& 2 \partial_{[\m} h_{\n]}^A  + f_{BC}{}^A h_\m^B h_\n^C \,,
 		\eea
 		we give the transformation rules for $e_\m{}^a$ and $\o_\m{}^{ab}$ and the corresponding curvatures $R_{\m\n}{}^a (P)$ and $R_{\m\n}{}^{ab}(M)$:
 		\bea
 		\d e_\m{}^a &=& \partial_\m \x{}^a + \o_\m{}^{ab} \x_b - \l^{ab} e_{\m b} \,,\nn\\
 		\d \o_\m{}^{ab} &=& \partial_\m \l^{ab} + 2 \o_{\m c}{}^{[a} \l^{b]c} \,.
 		\label{PoincareTransformationRules}
 		\eea
 		Here $\x^a$ and $\l^{ab}$ are the transformation parameters for $P_a$ and $M_{ab}$ respectively, and
 		\bea
 		R_{\m\n}^a (P) &=& 2 \partial_{[\m} e_{\n]}{}^a + 2 \o_{[\m}{}^{ab} e_{\n]b} \,,\nn\\ 
 		R_{\m\n}{}^{ab}(M) &=& 2 \partial_{[\m} \o_{\n]}{}^{ab} + 2 \o_{[\m}{}^{ac} \o_{\n]c}{}^{b} \,.
 		\label{PoincareCurvatures}
 		\eea
 		In order to identify $\o_{\m}{}^{ab}$ as the spin connection and make the vielbein $e_\m{}^a$ the only independent degree of freedom, we impose the curvature constraint
 		\bea
 		R_{\m\n}{}^a (P) &=& 0 \,,
 		\label{CurvatureCondition1}
 		\eea
 		which implies that $\o_{\m}{}^{ab}$ is given in terms of the vielbein as
 		\bea
 		\o_\m{}^{ab}(e) &=& 2 e^{\n [a} \partial_{[\m} e_{\n]}{}^{b]} - e^{\n[a} e^{b]\s} e_{\m c} \partial_\n e_\s{}^c \,.
 		\label{SpinConnection}
 		\eea
 		%and we define a Levi-Civita connection $\G_{\m\n}^\r$ by the structure equation $\nabla_\m e_\n{}^a = \partial_\m e_\n{}^a + \o_\m{}^a{}_b - \G_{\m\n}^\r e_\r{}^a = 0$.
 		
		With this identification, arbitrary contractions of the curvature of the spin-connection $R_{\m\n}{}^{ab}(M)$ given in \eqref{PoincareCurvatures} can be used to construct Poincar\'e invariant metric theories.
 		
 		When an additional local scale invariance is demanded, we can improve the Poincar\'e group with the scale symmetry generator $D$
 		\bea
 		P_a\,, \quad M_{ab} \,, \quad D \,,
 		\label{DilatationGenerators}
 		\eea
 		with the corresponding gauge fields
 		\bea
 		h_\m{}^A \equiv \{e_\m{}^a \,, \o_\m{}^{ab}\,, b_\m \} \,.
 		\label{DilatationGaugeFields}
 		\eea
 		In this case, the transformation rules and the curvatures are given by
 		\bea
 		\d e_\m{}^a &=& \partial_\m \x{}^a + \o_\m{}^{ab} \x_b - \l^{ab} e_{\m b} - \L_D 	e_\m{}^a \,,\nn\\
 		\d \o_\m{}^{ab} &=& \partial_\m \l^{ab} + 2 \o_{\m c}{}^{[a} \l^{b]c}  \,,\nn\\
 		\d b_\m &=& \partial_\m \L_D \,,
 		\label{ScaleInvariantTransformationRules}
 		\eea
 		where $\L_D$ is the parameter for scale transformations and 
 		\bea
 		R_{\m\n}{}^a (P) &=& 2 \partial_{[\m} e_{\n]}{}^a + 2 \o_{[\m}{}^{ab} e_{\n]b} + 2 b_{[\m} e_{\n]}{}^a \,,\nn\\
 		R_{\m\n}{}^{ab}(M) &=& 2\partial_{[\mu}\omega_{\nu]}{}^{ab}+2\,\omega_{[\mu}{}^{ac}\,\omega_{\nu]c}{}^{b}  \,,\nn\\
 		R_{\m\n} (D) &=& 2 \partial_{[\m} b_{\n]} \,.
 		\label{ScaleInvariantCurvatures}
 		\eea
 		Once again, in order to convert $P$-gauge transformations to general coordinate transformations, we impose the condition $R_{\m\n}{}^a (P) = 0$ and identify the field $\o_\m{}^{ab}$ as the spin connection. However, in this case, the definition of $\o_\m{}^{ab}$ picks up a $b_\m$ dependent term
 		\bea
 		\o_\m{}^{ab} (e,b) &=& \o_\m{}^{ab}(e)  + 2 e_\m{}^{[a} b^{b]} \,,
 		\label{ScaleSpinConnection}
 		\eea
 		where $\o_\m{}^{ab}(e)$ is as defined in (\ref{SpinConnection}). Furthermore, the Bianchi identity on $R_{\m\n}{}^a (P)$ now implies
 		\bea
 		e_{[\r}{}^a R_{\m\n]}(D) &=& R_{[\m\n\r]}{}^a (M) \,.
 		\label{BianchiOnP}
 		\eea
 		Although the arbitrary contractions of the curvature $R_{\m\n}{}^{ab}(M)$ are still Poincar\'e invariant, we now have to be careful with the scale invariance, i.e.,
 		\bea
 		\d_D R_{\m\n}{}^{ab} (M) &=& 0 \,,\nn\\
 		\d_D R_{\m}{}^a (M) &=& \L_D R_\m{}^a (M) \,,\nn\\
 		\d_D R(M) &=& 2 \L_D R(M) \,,
 		\eea
 		where we have defined $R_\m{}^a (M) \equiv e^\n{}_b R_{\m\n}{}^{ab}(M)$ and $R(M) \equiv  e^\m{}_a e^\n{}_b R_{\m\n}{}^{ab}(M)$. This implies that as $\sqrt{-g}$ transforms as $\d_D \sqrt{-g} = -4 \L_D \sqrt{-g}$, the covariant curvature $R(M)$ cannot itself define an invariant action unless we consider $R(M)^2$, or an additional scalar field $\f$ that transforms under scale transformations with the scaling dimension $\o \neq 0$ 
 		\bea
 		\d_D \f &=& \o \L_D \f \,.
 		\eea
 		Therefore, setting $\o = 1$ a scale invariant two-derivative gravity can be given as
 		\bea
 		S_{\text{R(M)}} &=& \int {\rm d}^4 x \, \sqrt{-g} \, \phi^2 \Big(R - 6 b_\m b^\m - 6 \nabla_\m b^\m \Big)\,, \qquad 
 		\label{ActionRM}
 		\eea
 		where $R$ is the standard Ricci scalar, and we have used the fact that 
 		\bea
 		R(M) = R - 6 b_\m b^\m - 6 \nabla_\m b^\m. 
 		\label{RM}
 		\eea
 		Similarly, we can introduce an action for the scalar field $\f$ as
 		\bea
 		S_\f &=& -\frac12 \int {\rm d}^4 x \, \sqrt{-g} \, \cD_a \f \cD^a \f \,,
 		\label{ActionScalePhi}
 		\eea
 		where the covariant derivative $\cD_a \f$ is defined by
 		\bea
 		\cD_a \f &=& e^\m{}_a ( \partial_\m \f - b_\m \f ) \,.
 		\label{CovariantDerivativePhi}
 		\eea
 		Finally, we can give an action for the vector field $b_\m$ as
 		\bea
 		S_{\text{b}} &=& -\frac14 \int {\rm d}^4 x \sqrt{-g} R_{\m\n}(D) R^{\m\n}(D)\,.
 		\label{bAction}
 		\eea
 		
 		When the symmetries are extended to the full conformal group, we include the  generator of the special conformal symmetry, thus, the generators are given by
 		\bea
 		P_a \,, \quad M_{ab}\,, \quad D\,, \quad K_a \,,
 		\eea
 		with the corresponding gauge fields
 		\bea
 		h_\m{}^A \equiv \{e_\m{}^a\,, \o_\m{}^{ab}\,, b_\m\,, f_\m{}^a\} \,.
 		\eea
 		In this case, the transformation rules and the curvatures are given by
 		\bea
 		\d e_\m{}^a &=& \partial_\m \xi^a + b_\m \xi^a + \o_\m{}^{ab} \xi_b - \l^{ab} e_{\m b} - \L_D e_\m{}^a \,,\nn\\
 		\d \o_\m{}^{ab} &=& \partial_\m \l^{ab} + 2 \o_{\m c}{}^{[a} \l^{b]c} - 4 \xi^{[a} f_\m{}^{b]} - 4 \L_{K}^{[a} e_\m{}^{b]} \,,\nn\\
 		\d f_\m{}^a &=& \partial_\m \L_K^a - b_\m \L_K^a  + \o_\m{}^{ab} \L_{K b} - \l^{ab} f_{\m b} + \L_D f_\m{}^a \,, \nn\\
 		\d b_\m &=& \partial_\m \L_D - 2 \xi^a f_{\m a} + 2 \L_K^a e_{\m a} \,,
 		\label{TransformationRulesConformal}
 		\eea 
 		where $\L_{Ka}$ is the parameter for special conformal transformation, and
 		\bea
 		\widehat R_{\m\n}{}^a (P) &=& 2 \partial_{[\m} e_{\n]}{}^a + 2 \o_{[\m}{}^{ab} e_{\n]b} + 2 b_{[\m} e_{\n]}{}^a \,,\nn\\
 		\widehat{R}_{\m\n}{}^{ab}(M) &=& 2\partial_{[\mu}\omega_{\nu]}{}^{ab}+2\,\omega_{[\mu}{}^{ac}\,\omega_{\nu]c}{}^{b}+8f_{[\mu}{}^{[a}e_{\nu]}{}^{b]} \,,\nn\\
 		\widehat R_{\m\n} (D) &=& 2 \partial_{[\m} b_{\n]} - 4 f_{[\m}{}^a e_{\n]a} \,,\nn\\
 		\widehat R_{\m\n}{}^{a} (K) &=& 2\partial_{[\m} f_{\n]}{}^a + 2 \o_{[\m}{}^{ab} f_{\n]b} - 2 b_{[\m} f_{\n]}{}^a \,,
 		\label{ConformalCurvatures}
 		\eea
 		where the hatted notation indicates the full covariance under conformal symmetries. We again need to use the condition $\widehat R_{\m\n}{}^a (P) = 0$ to convert $P$-gauge transformations to general coordinate transformations. However this time we also need to impose a second constraint to fix the gauge field of special conformal transformations,
 		\bea
 		e^\m{}_a\widehat  R_{\m\n}{}^{ab}(M) &=& 0 \,.
 		\label{CurvatureCondition2} 
 		\eea
 		As a result of these constraints, $\o_\m{}^{ab}$ and $f_\m{}^a$ become dependent fields. The definition of $\o_{\m}{}^{ab}$ remains the same as (\ref{ScaleSpinConnection}), and the definition of $f_\m{}^a$ is given by
 		\bea
 		f_\m{}^a (e,b) &=& -\tfrac{1}{4}\widehat{R}^\prime_{\mu}{}^{a}(M)+\tfrac{1}{24}e_{\mu}{}^{a}{\widehat{R}^\prime(M)}\ ,
 		\label{fmaDefinition}
 		\eea
 		where the prime indicates that $f_\m{}^a$ term is excluded in the definition of $\widehat R_{\m\n}{}^{ab}(M)$ in (\ref{ConformalCurvatures}).
 		
 		Unlike the Poincar\'e invariance and the scale invariance, these constraints and definitions have a dramatic consequence in the case of conformal symmetry: the only independent field that transforms under the special conformal transformation is $b_\m$ (\ref{TransformationRulesConformal}). Therefore, as the action has to be invariant under that symmetry, the $b_\m$ field cannot appear in a conformal theory of gravity. Therefore, although the scale symmetry is gauged and local, its gauge field does not show up in a conformally invariant gravity theory,
 		\bea
 		S_{\text{conf}} &=& \int {\rm d}^4 x \, \sqrt{-g} \Big( \frac1{12} \f^2 R - \frac12   \f \Box \f \Big) \,,
 		\label{ConformalAction}
 		\eea
 		where $\Box$ is the usual d'Alembertian operator. Note that the potential $\l \f^4$ can be added to both the conformal theory (\ref{ConformalAction}) and the scale invariant theory (\ref{ActionScalePhi}) as $\f$ is already invariant under special conformal transformation.
 		
 		%%%%%%%%%%%%%%%%%
 		\section{Conserved Current}
 		\label{ConservedCharge}
 		%%%%%%%%%%%%%%%%%
 		The fact that $b_\m$ does not show up in the conformal gravity models has a consequence that the conserved current associated with the scale symmetry vanishes \cite{Jackiw:2014koa,Hertzberg:2014aha}. This can be understood from the viewpoint that if a gauge field that gauges a certain symmetry does not show up in an action, then it is not possible to distinguish whether the action is invariant under the global or the local symmetry that is gauged by that field. For instance, the action (\ref{ConformalAction}) is invariant under the global transformations
 		\bea
 		\d g_{\m\n} = - 2 \L_D g_{\m\n} \,, \quad \d \f = \L_D \f  \,,
 		\label{LocalVariation}
 		\eea
 		and, as discussed, making the scale transformation parameter local, $\L_D = \L_D (x)$, does not introduce a vector field to the theory, and the model remains the same. Thus, one cannot distinguish the local and the global scale symmetries. As a result, if there were a non-vanishing conserved current for the global scale symmetry for this model, then there should also be a conserved current for the local scale symmetry. Hence, such symmetries should not have conserved currents. 
 		
 		In this section, our purpose is to verify this in two-derivative level explicitly. In order to do so, let us consider a scale invariant theory with a free parameter $\g$ such that when $\g$ takes a certain value, the scale invariant model reduces to the conformal theory (\ref{ConformalAction}). We let this model be
 		\bea
 		S &=& \int {\rm d}^4 x  \sqrt{-g}  \Big(  \tfrac1{12}\f^2 R(M) + \tfrac12 \g \cD_\m \f \cD^\m \f \nn\\
 		&& - \tfrac14 g^2 (1-\g) R_{\m\n} (D) R^{\m\n}(D) \Big) \,,
 		\label{PreScaleInv}
 		\eea
 		where we have combined the scale invariant actions (\ref{ActionRM}), (\ref{ActionScalePhi}), and (\ref{bAction}) with different coefficients. Here $g(1-\g)$ is a function of $\g$ that satisfies $g(0) = 0$. Note that when $\g = 1$, the kinetic term for the gauge field $b_\m$ drops out, and the $b_\m$ terms in the scale invariant Ricci scalar $R(M)$ and the kinetic terms for the scalar field $\cD_\m \f \cD^\m \f$ cancel each other out. To see that explicitly, we partially integrate the $b_\m \partial^\m \phi$ terms in the kinetic terms for the scalar field, and arrive to the following scale invariant model
 		\bea
 		S &=& \int \sqrt{-g}  \Big(  \tfrac1{12}\f^2 (R - 6 (1-\g) b_\m b^\m - 6 (1-\g) \nabla^\m b_\m ) \nn\\
 		&&  + \tfrac12 \g \partial_\m \f  \partial^\m \f   - \tfrac14 g^2 (1-\g)  R_{\m\n}(D)  R^{\m\n}(D) \Big) {\rm d}^4 x \,.\,\,
 		\label{ScaleInvariantTwoDerivative}
 		\eea
 		Thus, when $\g$ = 1, all the $b_\m$ terms drop out and we obtain the conformal action (\ref{ConformalAction}). Therefore, the free parameter $\gamma$ measures the deviation from conformal invariance while preserving the scale symmetry. 
 		%
 		%The field equations for $\f, b_\m$ and the metric $g_{\m\n}$ for the action (\ref{ScaleInvariantTwoDerivative}) are given by
 		%\bea
 		%0 &=& \frac16\f (R - 6(1-\a) b_\m b^\m - 6 (1-\a) \nabla^\m b_\m ) - \a \Box \f \,,\nn\\
 		%0 &=& (1-\a) (4\nabla^\n F_{\m\n} + \f^2 b_\m - \f \partial_\m \f) \,,\nn\\
 		%0 &=& \tfrac1{12} \f^2 G_{\m\n} - \tfrac1{12} (\nabla_\m \nabla_\n - g_{\m\n} \Box)\f^2 \nn\\
 		%&& +\tfrac14 (1-\a)g_{\m\n} \f^2 (b_\l b^\l + \nabla_\l b^\l) \nn\\
 		%&& - \tfrac12 (1-\a) \f^2 (b_\m b_\n + \nabla_\m b_\n) -\tfrac14 \a g_{\m\n} \partial_\l \f \partial^\l \f \nn\\
 		%&&  + \tfrac12 \a \partial_\m \f \partial_\n \f - \frac12 (1-\a) g_{\m\n} F_{\r\s} F^{\r\s} \nn\\
 		%&& + 2 (1-\a) F_{\m}{}^\s F_{\n\s} \,.
 		%\eea
 		
 		The Lagrangian in (\ref{ScaleInvariantTwoDerivative}) is a scalar with scaling dimension $\o = 0$ by construction, thus $\d_D \cL = 0$, not contributing to the conserved current via a surface term. Hence, the conserved current is given by \cite{Oda:2013uca,Oda:2016pok}
 		\bea
 		j^\m_D &=& - \frac{\partial\cL}{\partial (\partial_\m \f)} \d \f - \frac{\partial \cL}{\partial (\partial_\m g_{\n\r})} \d g_{\n\r} \,.
 		%&=& 2 \L g_{\n\r} \frac{\partial \cL}{\partial (\partial_\m g_{\n\r})} -\L\f   \frac{\partial\cL}{\partial (\partial_\m \f)} 
 		\label{ConservedCurrentFormulae}
 		\eea
 		%Varying the action with respect to $\partial_\m \f$, we obtain
 		%\bea
 		%\frac{\partial\cL}{\partial (\partial_\m \f)} &=& \sqrt{-g} g^{\m\n} ( \f b_\n + \a (\partial_\n - b_\n) \f ) \,,
 		%\eea
 		%and the variation with respect to $\partial_\m g_{\n\r}$, we have
 		%\bea
 		% \frac{\partial \cL}{\partial (\partial_\m g_{\n\r})} &=& - \sqrt{-g} \f \partial_\m \f 
 		%\eea
 		
		Using the transformations (\ref{LocalVariation}), the conserved current is 
 		\bea
 		j^\m_D &=& (1-\g) \sqrt{-g} g^{\m\n} (\partial_\n - 2 b_\n) \f^2 \,.
 		\label{ConservedCurrent}
 		\eea
 		As expected, the current vanishes when $\g = 1$. Furthermore, if $b_\m$ is not dynamical, which corresponds to $g(1-\g) = 0$, then the field equation for $b_\m$ simply takes the model back to the conformal theory (\ref{ConformalAction}), not providing a conserved current. Thus, to evade the ``fake gauge invariance" \cite{Jackiw:2014koa,Hertzberg:2014aha}, one needs to set $\g \neq 1$ and dynamically gauge the scale symmetry, which cannot be done in the case of conformal theories due to special conformal invariance. 
 		
Now that we have ensured the parameter $\g$ must satisfy $\g \neq 1$, which guarantees that $g(1-\g) \neq 0$, we can perform a field redefinition in a way that the coefficient of the kinetic term $R_{\m\n}(D)  R^{\m\n}(D)$ takes its canonical form properly at the expense of changing the coefficient of the kinetic term of the scalar field. This can be done by the following field redefinition of the gauge fields, the gauge parameter $\L_D$ and the scalar field:
		  		\bea
 		&& b_\m = \frac{\bar{b}_\m}{g} \,,\,\,  \f = \bar{\f}^{1/g} \,,\,\, e_\m{}^a = \bar{\f}^{\b} \bar{e}_\m{}^a \,, \,\,  \L_D = \frac{\bar{\L}_D}{g}\,,
 		\eea
 		where $\b \equiv 1- 1/g$. Accordingly, we find that the covariant derivative of the scalar field, the spin connection, and the scale invariant Ricci tensor become
 		\bea
 		\cD_\m \f &=& \tfrac{1}{g} \bar{\f}^{-\b} \cD_\m \bar{\f} \,,\nn\\
 		{\o}_{\m}{}^{ab} (e,b) &=& \bar{\o}_{\m}{}^{ab} (\bar{e},\bar{b}) + 2 \b \bar{\f}^{-1} \bar{e}_\m{}^{[a} \cD^{b]} \bar{\f} \,,\nn\\
 		R(M) &=& \bar{\f}^{-2\b} \Big( \bar{R}(M) - 6 \b^2 \bar{\f}^{-2} \cD_\m \bar\f \cD^\m \bar\f \nn\\
 		&& - 6 \b \cD^\m (\bar{\f}^{-1} \cD_\m \bar\f)\Big) \,, 
 		\eea
 		where $\bar{R}(M) = \bar{R} - 6 \bar{b}_\m \bar{b}^\m - 6 \nabla^\m \bar{b}_\m$. As a consequence, the action given in (\ref{PreScaleInv}) can be equivalently written as 
		\bea
 		S &=& \int \sqrt{-\bar{g}} \Big(\tfrac{1}{12} \bar{\f}^2 \bar{R}(M) + \tfrac{1}{2} (\tfrac{\g-1}{g^2} +1) \cD_\m \bar\f \cD^\m \bar\f  \nn\\
 		&& - \tfrac{1}{4} \bar{R}_{\m\n} (D) \bar{R}^{\m\n} (D) \Big) {\rm d}^4x  \,.
		\label{1}
 		\eea
 		Note that in this new basis, the deviation from conformal invariance while preserving the scale symmetry is controlled by a combination of $\g$ and $g$, which we define as a new constant $\a$
 		\bea
 		\a \equiv \frac{\g-1}{g^2} +1\,.
 		\eea
 		After a partial integration of the $\bar b_\m \partial^\m \bar \phi$ terms in (\ref{1}), we find that the action (\ref{ScaleInvariantTwoDerivative}) can also be equivalently written as
 		\bea
 		S &=& \int \sqrt{-\bar g}  \Big(  \tfrac1{12}\bar\f^2 (\bar R - 6 (1-\a) \bar  b_\m \bar  b^\m - 6 (1-\a) \nabla^\m \bar  b_\m ) \nn\\
 		&&  + \tfrac12 \a \partial_\m \bar \f  \partial^\m \bar \f   - \tfrac14   \bar R_{\m\n}(D)  \bar R^{\m\n}(D) \Big) {\rm d}^4 x\,\,
 \,,
		\label{ScaleIncFin}
 		\eea
 		which is invariant under the following scale transformations
 		\bea
 		\d \bar{g}_{\m\n} = -2 \bar{\L}_D \bar{g}_{\m\n}  \,, \quad \d \bar{\f} = \bar{\L}_D \bar{\f} \,, \quad \d \bar{b}_\m = \partial_\m \bar{\L} \,.
 		\eea
		Thus we arrived at the action \eqref{ScaleIncFin} that we can consider for working on the inflationary dynamics based on the spontaneous breaking of the scale symmetry. As will be demonstrated in the next section, the constant $\a$ here is the parameter that characterizes the inflaton field as in the superconformal $\a$-attractors. In the scale invariant setting, as $\a$ enters to the vector potential in a certain way, it also determines the dynamics of the vector field, which, together with the scalar field (inflaton) characterizes the evolution of the universe in our model. Therefore the parameter $\a$ is constrained by the dynamics of both the scalar and the vector field in the scale invariant setting of $\a$-attractors.
		
		 		%%%%%%%%%%%%%%%%%
 		
 		%%%%%%%%%%%%%%%%%
 		\section{Broken Scale Invariance and $\a$-attractors}
 		\label{AlphaAttractors}
 		%%%%%%%%%%%%%%%%%
 		
 		As we have paved the way for the scale invariance in the previous sections, we can now proceed to work on the consequences of a dynamically gauged vector field in the context of inflationary cosmology. Following \cite{Kallosh:2013hoa}, we first write down the dynamically gauged $SO(1,1)$ and scale invariant theory with two scalar field $\f$ and $\chi$ of scaling dimension $\o = 1$ (see \cite{Ozkan:2015kma}), which is based on the action (\ref{ScaleIncFin})
 		\bea
 		S &=&  \int {\rm d}^4 x \sqrt{-g} \Big( \tfrac1{12}(\chi^2 - \f^2) (R - 6 (1-\a) b_\m b^\m  \nn\\
 		&& - 6 (1-\a) \nabla^\m b_\m )  + \tfrac12 \a \partial_\m \chi \partial^\m \chi - \tfrac12 \a \partial_\m \f \partial^\m \f  \nn\\
 		&& - \tfrac14 R_{\m\n} (D) R^{\m\n} (D)   - \tfrac1{36} f(\frac{\f}{\chi}) (\chi^2 - \f^2)^2 \Big) \,.
 		\label{ScaleAction}
 		\eea
 		Note that we have dropped the bar in the definition of the gauge fields and the scalar fields, as we will work in this basis in the rest of the paper. In order to obtain a Poincar\'e gravity from the two-scalar model (\ref{ScaleAction}), we impose
 		the gauge fixing condition $\chi^2 - \phi^2 = 6 M^2_{\rm pl}$, which sets $\L_D = 0$. This condition has the solution
 		\bea
 		\chi = \sqrt6 M_{\rm pl} \cosh \frac{\vf}{\sqrt{6  \a}M_{\rm pl}} ,  \nn\\ \quad  \f = \sqrt6 M_{\rm pl} \sinh \frac{\vf}{\sqrt{6  \a}M_{\rm pl}} \,,
 		\eea
 		in terms of a canonical scalar $\vf$. As a result of the gauge fixing, the model reduces to
 		\bea
 		S &=& \int {\rm d}^4 x \sqrt{-g} \Big( \tfrac{1}{2}M_{\rm pl}^2 R  -\tfrac14 R_{\m\n} (D) R^{\m\n} (D) - \tfrac12 \partial_\m   \vf \partial^\m \vf \nn\\
 		&&   - M_{\rm pl}^4 f(\tanh \frac{\vf}{\sqrt{6\a} M_{\rm pl}}) - 3 (1-\a) M_{\rm pl}^2 b_\m b^\m \Big)\,,
 		\label{Model}
 		\eea
 		where we have dropped the $\nabla^\m b_\m$ term as it is a total derivative. When the vector field $b_\m$ lacking, this model is given by the bosonic part of the superconformal $\a$-attractors \cite{Kallosh:2013yoa}. In this case, for $\a=1$, the scalar potential gives rise to universal predictions for the inflationary observables \cite{Kallosh:2013hoa}
 		\bea
 		n_s = 1 - \frac2N, \qquad  r = \frac{12}{N^2} \,,
 		\label{a1nsr}
 		\eea
 		which coincides with the predictions of a large class of inflationary models including Starobinsky inflation \cite{Starobinsky:1980te}, Higgs inflation \cite{Salopek:1988qh,Bezrukov08}, the non-minimal inflationary models $\x\f^2R$ with negative non-minimal coupling $\x\leq-10^{-1}$ \cite{Kallosh:2013maa}, and generalised non-minimally coupled model $\x \sqrt{V(\f)} R$ \cite{Kallosh:2013tua} at large non-minimal coupling. 
 		If $\a \lesssim 1$, and $N \gg 1$, this model leads to universal predictions \cite{Kallosh:2013yoa}
 		\bea
 		n_s = 1 - \frac2N, \qquad  r = \frac{12 \a}{N^2} \,.
 		\label{ansr}
 		\eea
 		When $\a$ is large, one may encounter a wide variety of possible scenarios, e.g., when the simplest class of models $V(\vf) = \tanh^{2n}(\vf/\sqrt{6\a}M_{\rm pl})$ corresponding to super-conformal $\a$-attractors are considered, the large $\a$ limit leads to the predictions of the chaotic inflation with $V(\f) \sim \f^{2n}$ \cite{Kallosh:2013yoa}
 		\bea
 		n_s = 1 - \frac{2n+2}{2N+n}, \qquad  r = \frac{16n}{2N+n} \,.
 		\label{LargeAlpha}
 		\eea
 		In the opposite limit, $\a \rightarrow 0$, there is a universal attractor prediction
 		\bea
 		n_s = 1 - \frac2N, \qquad  r = 0 \,.
 		\eea
 		There are also models that correspond to the values of $\a$ in between, e.g. $\a = 1/9$ corresponds to 
 		Goncharov-Linde model \cite{Goncharov:1983mw,Goncharov:1984,Linde:2014hfa} with $r \approx 0.0003$. The stability of the supersymmetric realization of the $\a$-attractors, including the $\a \rightarrow 0$ limit are discussed in detail in \cite{DiederikMarco,Marco,Carrasco:2015pla,Kallosh:2013yoa}. Nevertheless, for a purely bosonic model, there is no reason, or a mechanism to prefer any value of $\a$. 
 		On the other hand, not all values of $\a$ are experimentally favoured, such that the latest upper bound given by Planck $r<0.11$ (95\% CL) leads to $\a<7.1$ (95\% CL) for superconformal $\a$-attractors \cite{Kallosh:2013yoa} and $\a<14.1$ (95\% CL) for $\a$-attractors motivated by considering a vector for the inflaton in supergravity \cite{Ferrara:2013,Kallosh:2013yoa}, namely, when $\a$ increases, $r$ grows 
 		(\ref{ansr}) and one moves to the disfavored regions of the Planck data \cite{Ade:2015lrj}. As we shall see in the following section, in contrast to the purely bosonic model, the presence of the vector field $b_\m$ can restrict $\a$, as it can determine the vector field mass, which in turn can confine the preferable values of $\a$ to a range much narrower than the rather large range $\alpha\lesssim10$ allowed by the observational data \cite{Ade:2015lrj}.
 		%%%%%%%%%%%%%%%%%
 		
 		%%%%%%%%%%%%%%%%%
 		\section{Dynamics of the massive vector field and Restrictions on $\alpha$}
 		\label{VectorFieldStability}
 		%%%%%%%%%%%%%%%%%
 		To investigate the dynamics of the vector field, we choose, without loss of generality, $f(\f/\chi) = \f^2 / (\f+\chi)^2$, that leads to
 		\bea
 		f(\tanh \frac{\vf}{\sqrt{6\a} M_{\rm pl}}) = \frac34 \frac{M^2}{M_{\rm pl}^2} \Big(1-e^{-\sqrt{\frac{2}{3\a}}\frac{\vf}{M_{\rm pl}}}\Big)^2 \,,
 		\label{Potential}
 		\eea 
 		where $M$ is some mass scale. With this choice, the field equations for the metric $g_{\m\n}$, the scalar field $\vf$ and the vector $b_\m$ are given by 
 		\bea
 		\label{metricequation}
 		0 &=& G_{\m\n}-  \tfrac{1}{M_{\rm pl}^2} \Big(T_{\m\n}^\vf + T_{\m\n}^b \Big)\,, \\
 		0 &=& \Box \vf  + \sqrt{\tfrac{3}{2\a}} M_{\rm pl} M^2 e^{-\sqrt{\frac{2}{3\a}}\tfrac{\vf}{M_{\rm pl}}} \Big(1 - e^{-\sqrt{\frac{2}{3\a}}\tfrac{\vf}{M_{\rm pl}}} \Big) \,,\qquad \\
 		0 &=& \nabla_\m R^{\m\n} (D) - 6  (1-\a) M_{\rm pl}^2 \, b^\n \,.
 		\label{EquationOfMotion}
 		\eea
 		Here $T_{\m\n}^b$ and $T_{\m\n}^\vf$ are the energy-momentum tensors of the minimally coupled scalar and vector fields, respectively, and read
 		\bea
 		\label{scalarEMT}
 		T_{\m\n}^\vf &=& \partial_\m \vf \partial_\n \vf - \frac12 g_{\m\n} \partial_\l \vf \partial^\l \vf\nn\\
 		&&  - \tfrac34  g_{\m\n} M^2 M_{\rm pl}^2 \Big(1 - e^{-\sqrt{\frac{2}{3\a}}\frac{\vf}{M_{\rm pl}}} \Big)^2  \,,\\
 		\label{vectorEMT}
 		T_{\m\n}^b &=&   R_\m{}^\r (D) R_{\n\r}(D) -  \tfrac14 g_{\m\n} R_{\r\s} (D) R^{\r\s} (D) \nn\\
 		&& - 3 g_{\m\n} (1-\a) M_{\rm pl}^2 b_\l b^\l  + 6 (1-\a) M_{\rm pl}^2 b_\m b_\n  \,.
 		\eea
 		
 		The field equations for the scalar and the vector are completely decoupled; therefore the vector field does not affect the well-known inflationary dynamics of the scalar field.
 		Here, we are interested in the consequences of the intervention of the vector field in the model, which can yield anisotropic pressure. Hence, let us consider the simplest spacetime metric that can accommodate anisotropic pressure, namely
 		%the spatially flat and homogeneous but not necessarily isotropic locally rotationally symmetric 
 		(LRS) Bianchi type-I spacetime metric,
 		\bea
 		ds^2 &=& - dt^2 + c^2 (t) [dx^2  + dy^2] + a^2 (t) dz^2 \,.
 		\label{Metric}
 		\eea
 		With this metric, we first observe that $G_{0i} = 0$ indicating \cite{Koivisto:2008xf}
 		\bea
 		T_{0i} &=& 6 (1-\a)b_0 b_i = 0 \,.
 		\eea
 		As we consider $\a \neq 1$, one must make a choice: either the vector $b_\m$ is timelike $b_\m = (\s, 0)$, or spacelike $b_\m = (0, b_i)$. If the vector is chosen to be timelike, the vector field equation would trivially be  satisfied. Therefore, we focus on the spacelike case and assume a homogeneous vector field which lies in the $z$-direction
 		\bea
 		b_\m (t) = b_z (t) \,,
 		\eea
 		which is consistent with our choice of metric \eqref{Metric}. With this choice, the energy density and pressures for the vector field read
 		\bea
 		\label{vectorenergydensity}
 		\r^b &=& \frac{\dot{b}_z^2}{2a^2} + m^2\, \frac{b_z^2}{2a^2} \,,\\
 		p_x^b = p_y^b &=& \frac{\dot{b}_z^2}{2a^2} - m^2\, \frac{b_z^2}{2a^2} \,,\\
 		p_z^b &=& -\frac{\dot{b}_z^2}{2a^2} + m^2\, \frac{b_z^2}{2a^2} \, ,
 		\eea
 		where $p_x^b$, $p_y^b$ and $p_z^b$ are the pressures along the $x$-, $y$- and $z$-axes. Here
 		\bea
 		\label{VectorMass}
 		m^2=6 (1-\a) M_{\rm pl}^2
 		\eea
 		is the mass-squared of the vector field, which can easily be read from \eqref{EquationOfMotion} or \eqref{vectorEMT}. We note that the vector field yields an $m$-dependent anisotropic equation of state (EoS) parameter;
 		\bea
 		w_x^b = w_y^b = - w_z^b = \frac{\dot{b_z^2} - m^2 b_z^2}{\dot{b_z^2} + m^2 b_z^2} \,,
 		\eea
 		where $w_x^b=p_x^b/\rho^b$, $w_y^b=p_y^b/\rho^b$ and $w_z^b=p_z^b/\rho^b$ are the directional EoS parameters along the $x$-, $y$- and $z$-axes.
 		
 		These two equations are important for investigating the consequences of the presence of a massive vector field in our model. First of all, we see from \eqref{VectorMass} that $\alpha$ cannot take arbitrary values but rather is restricted to take values in the following range
 		\bea
 		\frac56 < \a < 1 \,.
 		\eea
 		The upper bound is required to keep mass-squared of the vector field positive $m^2>0$, since an imaginary (tachyonic) mass leads to a ghost instability \cite{Himmetoglu09a,Himmetoglu09b}, and the lower bound exists to keep the mass lower than $M_{\rm pl}$. In contrast to the $\a$-attractors from a broken (super)conformal symmetry, in which $\a$ can take arbitrary positive values, the broken scale invariance can severely restrict $\a$ as it depends on the mass of the vector field. This result is somewhat unexpected as the usual dictum is that more symmetry implies higher constraint. Accordingly, it turns out that the inflationary observables of the model are
 		\bea
 		n_s=1-\frac{2}{N},\quad r=\frac{12}{N^2} - \frac{2}{N^2} \frac{m^2}{M_{\rm pl}^2},
 		\eea
 		where $0<m^2<M_{\rm Pl}^2$ implying
 		\bea
 		\frac{10}{N^2}<r<\frac{12}{N^2},
 		\eea 
 		where this range is at the sweet spot of the Planck data \cite{Ade:2015lrj}.
 		
 		The range of $\a$ should be further investigated considering not only the inflationary but also the post-inflationary era. First, we should ensure that the vector field yielding an $m$-dependent anisotropic EoS does neither alter the inflationary dynamics, nor avoid isotropization during inflation. We should also make sure that the vector field does not considerably alter the post-inflationary dynamics; the isotropy of the space should be maintained or any possible anisotropization during this stage should end and be compensated before the Big Bang Nucleosynthesis (BBN) takes place. As dictated by the cosmic no-hair theorem, the vector field will be dominated by the scalar field during the inflationary era, namely, when the slow roll conditions are satisfied, and hence the universe will isotropize and evolve towards the de Sitter solution on an exponentially rapid time scale \cite{nohair1,nohair2,nohair3,nohair4}. Therefore, whether it yields heavy or light mass, the vector field does not alter the dynamics of the universe during the inflationary epoch. Nevertheless its dynamics during inflation is still important for the post-inflationary era and should be discussed. To do so, let us turn attention to the vector field equation
 		\bea
 		\ddot{b}_z + (2H_c - H_a) \dot{b}_z  + m^2 b_z &=& 0 \,,
 		\label{VectorEOM2}
 		\eea
 		where $H_a \equiv \dot{a}/a$ and $H_c \equiv \dot{c}/c$ are the directional Hubble parameters. Relying on the cosmic no-hair theorem we can take $H_a\simeq H_c \approx {\rm constant}$, which is typically about $ H_{\rm inflation}\sim10^{-5}\,M_{\rm pl}$ during inflationary era. Let us first assume that the mass of the vector field is very heavy $m\sim M_{\rm pl}$ implying $m \gg H_{\rm inflation}$. In this case the mass term in \eqref{VectorEOM2} dominates over the friction term, and the vector field engages in an underdamped harmonic oscillations around $b_z = 0$ with an envelope decreasing as $\propto v^{-1/2}$, where $v$ is the mean scale factor defined as $v=({ac^2})^{1/3}$ (see \cite{Dimopoulos:2006ms}). Therefore, such a vector field would decay during inflation and cannot survive to dominate the universe after inflation. Consequently, a very heavy vector field ($m\sim M_{\rm pl}$) does never spoil the isotropy of the universe, and hence preserves the universal predictions of the $\a$-attractors, but coming with a bonus that predicts an almost exact value for $\alpha$, namely, $\alpha\approx5/6$ implying $r\approx 10/N^2$. Let us next assume that the vector is very light $m\ll H_{\rm inflation}$, i.e. $\alpha\approx 1$. In this case, the friction term in the vector equation (\ref{VectorEOM2}) dominates over the mass term and, thus, the vector field becomes overdamped and remains frozen, $\dot{b}_z = 0$, during inflation. This implies that if the vector field is light, it can survive inflation and can dominate the universe after that period. Hence, if the vector field has a mass $m<H_{\rm inflation}$, then we should further investigate its dynamics during the post-inflationary era. As long as $m<H$ during the post-inflationary era the vector field keeps on being overdamped and frozen so that it yields anisotropic EoS as $w_x =w_y = -w_z =-1$ and its energy density decreases as $\rho^b\propto a^{-2} \sim v^{-2}$ provided that the expansion is not highly anisotropic. According to this, the energy density of the vector field decreases slower than that of the radiation $\rho_{\rm rad}\propto v^{-4}$. Therefore, it would dominate over radiation and lead to a sufficiently large anisotropic expansion  due to its anisotropic EoS, which would spoil the BBN. This issue can be resolved in two different ways. (i) Let us first consider $m\ll H_{\rm BBN}\ll H_{\rm inflation}$, where $H_{\rm BBN}\sim 10^{-21}\,M_{\rm pl}$ is the Hubble parameter when BBN approximately takes place. In this case, we should assume that the value of $\rho^b$ at the end of inflation is so small that it stays subdominant until the end of BBN. However, if it has not decayed yet at the end of this period, it would still cause an anisotropization at a period after the BBN processes take place. (ii) Let us now consider $m\gg H_{\rm BBN}$ though $m<H_{\rm inflation}$. Since the average Hubble parameter $H\equiv(2H_c+H_a)/3$ decreases with the cosmic time during the post-inflationary era, a vector field with $m<H_{\rm inflation}$ can become $m\gg H$ at some point before BBN starts. If this can be achieved sufficiently long before the BBN period, then we can evade spoiling BBN due to anisotropization. The reason is that, once the condition $m\gg H$ is achieved, the vector field is no more light compared to the Hubble parameter and it would engage in an underdamped harmonic oscillations around $b_z = 0$. In this case, it would yield an EoS parameter oscillating about zero, which results in an isotropic EoS parameter equal to zero on average, $w_x =w_y = -w_z =0$ (see \cite{Dimopoulos:2006ms}). Hence it would behave like isotropic pressure-less matter with $\rho^b\propto v^{-3}$ and would not cause anisotropic expansion anymore. In this case, the universe, which was highly isotropized during the inflationary era, would undergo a temporary and moderate anisotropization process that can end and be compensated before the BBN takes place. Afterwards, since the value of $H$ has always been decreasing from the end of inflation until today, the condition $m\gg H$ would always be satisfied once the condition $m\gg H_{\rm BBN}$ is imposed, therefore an anisotropization due to vector field would not be an issue anymore.
 		
 		Thus we arrive at two characteristic types of scenarios that are in line with the standard BBN and $\Lambda$CDM models:
 		
 		\begin{itemize}
 			\item
 			The case $M_{\rm pl}>m\gg H_{\rm inflation}$ preserves the universal predictions of the $\alpha$-attractors and predicts a range $10^{-10}\ll 6 \; (1-\alpha)< 1$ implying $10^{-10}\ll 12-r\,N^2< 2$. This implies and leads to an isotropic universe.
 			\item
 			The case $H_{\rm BBN}\ll m\ll H_{\rm inflation}$ preserves the inflationary dynamics given by the $\alpha$-attractors, predicts $10^{-42}\ll 6(1-\alpha) \ll 10^{-10}$ and implies $10^{-42}\ll 12-r\,N^2 \ll 10^{-10}$. In this case, our model makes an additional prediction that the universe enters into a temporary anisotropization epoch just after the end of inflation. However, this epoch can end sufficiently long before the beginning of BBN and expansion anisotropy can stay at moderate values such that it can drop down to the values that do not spoil the standard BBN model once again.
 			
 		\end{itemize}

 		\section{Conclusions}
 		\label{Conclusions}
 		%%%%%%%%%%%%%%%%%
 		
 		In this paper, we have examined the features of the $\a$-attractor scenario if it arises from the spontaneous breaking of a scale symmetry. The key difference between the scale invariant scenario presented in this paper and the conformal setting \cite{Kallosh:2013hoa} is that our model includes a dynamical vector field, which can lead to non-trivial constraints on the dynamics of inflation.
 		
 		First of all, as opposed to the conformal model, the conserved current that is associated with the global scale symmetry of the model is non-vanishing. Second, although the scalar potentials of the conformal attractors and the scale invariant model precisely coincide, the dynamical vector field in our model \eqref{ScaleIncFin} severely constrains the $\a$-parameter to satisfy $5/6 < \a < 1$, and these values of $\a$ are consistent with latest Planck data \cite{Ade:2015lrj} on the inflationary observables $n_s$ and $r$ .
 		
 		This narrow range for the $\a$ that the mass of the vector field provides does not make a tangible difference between the different values for the inflationary observables and it might not be possible to distinguish it observationally. However, this narrow range of $\a$ covers the entire mass spectrum of the vector field, i.e. $0 < m < M_{\rm pl}$. Thereby, even the tiny changes in $\a$ leads to huge changes in the mass of the vector field whose magnitude, when compered with the value of the Hubble parameter, leads it to behave as either isotropic or anisotropic fluid.  Consequently, the mass of the vector field, which is determined by the $\a$ parameter, plays the main role in controlling the expansion anisotropy. Thus, different values of $\a$ cause dramatically different post-inflationary behaviors. Here, we presented two characteristic types of scenarios that are in line with the standard BBN and $\L$CDM: (i) $M_{\rm pl} > m \gg  H_{\text{inflation}}$, which preserves the universal predictions of the $\a$ attractors, and (ii)  $H_{\rm BBN}\ll m\ll H_{\rm inflation}$, which preserves the inflationary dynamics given by the $\alpha$-attractors, but which additionally predicts a temporary anisotropization epoch between the end of the inflation and the BBN.
 		
 		There are numerous directions that one can consider to proceed in this study. First of all, here we discuss the post-inflationary dynamics of the universe when the mass of the vector field satisfies $M_{\rm pl} > m \gg M_{\text{inflation}}$ or $H_{\rm BBN}\ll m\ll H_{\rm inflation}$. A thorough analysis that covers the entire region of the mass spectrum remains as an open problem. Moreover, the temporary anisotropization epoch, which occurs when  $H_{\rm BBN}\ll m\ll H_{\rm inflation}$, deserves further study.  An intriguing problem is whether the vector field that appears as a consequence of the gauged scale symmetry can be used to produce curvature perturbations as in the curvaton scenario \cite{curv1,curv2,curv3,curv4}. In this paper, we have not exhausted all the possibilities that the scale invariance can dictate in the action (\ref{ScaleAction}). Indeed, one can multiply the curvature of the vector field with a weight zero function of the scalars,
 		\bea
 		g({\f /\chi}) R_{\m\n}(D) R^{\m\n} (D) \,,
 		\eea
 		which preserves the scale invariance of the action (\ref{ScaleAction}). Then, by an appropriate choice of the function $g(\f/\chi)$, one can generate a scenario such that the kinetic term of the vector field flips signature before and after inflation, which successfully realizes the vector curvaton scenario \cite{Dimopoulos:2006ms}, and unifies the vector curvaton scenario with $\a$-attractors. 
 		
 		The scale invariant model that is introduced in this paper is somewhat a SUGRA-inspired model since $\cN=1$ SUSY has a $U(1)$ R-symmetry, therefore the local supersymmetry includes a vector field $A_\m$ in its spectrum. This vector field is auxiliary and is eliminated by its field equation in the 2-derivative theory \cite{Freedman:2012zz}. However, from a field theoretical viewpoint, the 2-derivative theory cannot be the full story and one has to take higher-order effects into account. When higher derivative terms show up, their supersymmetric completion includes kinetic terms for the auxiliary vector field, thus the $U(1)$ R-symmetry becomes dynamical, see e.g. \cite{deRoo:1990zm}; then the vector field cannot be simply integrated out. Therefore, we believe that the analysis we performed here for the dynamical vector field will be helpful to uncover the properties of supersymmetric $\a$-attractors in the presence of higher order corrections, which is an interesting direction to pursue.
		
		It has been shown that one might get logarithmic enhancements due to loop corrections during inflation  \cite{weinberg1,weinberg2}. Therefore one could consider including those effects for the present work. But there are two things that we need to keep in mind before doing that. The first one is the fact that conformal invariance decreases the emergence rate of virtual particles, which is the source of the loop effects \cite{woodard}. Because of this, loop contributions due to conformally coupled scalar particles are doomed to be very small. Hence, the dominant contribution would have to come from gravitons running in the loops.  The second fact is that quantum gravitational effects suffer from smallness of the loop counting parameter $G \, H^2$, where $G$ is Newton's constant and $H$ is the Hubble rate at the time of horizon exit. Taking these two points into account, one could include the effect of gravitons on scalar particles and hope to get a logarithmic enhancement which might counterbalance the smallness of the loop counting parameter as a future direction.
		
		Finally, the procedure introduced here to construct a scale-invariant generalization of the alpha-attractors can be applied to more wide examples of attractors potentials, such as the ones introduced in \cite{Dimopoulos:2017zvq, Odintsov:2016jwr}. This would give rise to a richer class of potentials with different constraints from the dynamics of the vector field, which would be an important step towards more realistic scenarios for inflationary attractors with a dynamical vector field.
 		
 		%%%%%%%%%%%%%%%%%
 		
 		%%%%%%%%%%%%%%%%%
 		\section*{Acknowledgements}
 		%%%%%%%%%%%%%%%%%
 		
 		We thank Gokhan Alkac, Glenn Barnich and Shahin Sheikh-Jabbari for valuable comments on conserved charges, and Renata Kallosh, Andrei Linde, Diederik Roest and Marco Scalisi for valuable comments on superconformal $\a$-attractors. \"{O}A acknowledges the support by the Distinguished Young Scientist Award BAGEP of the Science Academy. EOK acknowledges the support by Turkish Academy of Sciences Distinguished Young Scientist Award  T\"{U}BA-GEB\.{I}P 2015. The work of MO is supported in part by Marie Curie Cofund (No.116C028).

 	\end{document}